\documentclass[preprint2]{emulateapj}
\newcommand{\bb}{\begin{equation}}
\newcommand{\ee}{\end{equation}}

\shorttitle{Sunspot seismology}
\shortauthors{Rajaguru}

\begin{document}

\title{Separating surface magnetic effects in sunspot seismology: new time-distance helioseismic diagnostics} 
\author{S.P. Rajaguru\altaffilmark{1}} 
\affil{W.W. Hansen Experimental Physics Laboratory, Stanford University, Stanford CA 94305} 
\email{rajaguru@sun.stanford.edu}    
\altaffiltext{1}{Currently at: Indian Institute of Astrophysics, Bangalore - 560034, India}    

\begin{abstract} 
Time-distance helioseismic measurements in surface- and deep-focus geometries
for wave-paths that distinguish surface magnetic contributions from those 
due to deeper perturbations beneath a large sunspot are presented and analysed.
Travel times showing an increased wave speed region extending down to about 18 Mm beneath
the spot are detected in deep-focus geometry that largely avoids use of wave field 
within the spot. Direction (in- or out-going wave) and surface magnetic field (or focus depth) dependent
changes in frequency dependence of travel times are shown and identified to be signatures
of wave absorption and conversion in near surface layers rather than that of shallowness
of sunspot induced perturbations.
\end{abstract}

\section{Introduction}
\label{sec:intro}

A problem of major importance in solar and stellar magnetohydrodynamics is
the determination of subsurface magnetic and thermal constitution of
sunspots \citep{thomasweissbook}. Application of time-distance helioseismology
\citep{duvalletal93} in 3-d tomographic imaging of subsurface layers of
sunspots \citep{duvalletal96,sashaduvall99,zhaoetal01,zhaoetal06} provided
a major milestone in the seismology of sunspots \citep{thomasetal82,thomasetal88}.
Continuing developments in local helioseismology, 
notably helioseismic holography accompanied by theoretical modelling, have since 
identified several "surface" contributors to seismic measures, collectively known as
"surface magnetic effects" \citep{braun97,schunkeretal05,lindseyandbraun05a,lindseyandbraun05b,
zhaoetal06}, that arise from interactions of p modes with 
surface magnetic field of a sunspot. However, such interactions have so far not been
explicitly included in helioseismic inversions, for lack of a suitable model, 
and hence relative contributions of surface and sub-surface perturbations
have so far not been estimated.

Use of oscillation signals observed within sunspots
is known to be the primary source of most surface effects in sunspot seismology
\citep{braun97,lindsey06}, as well as the associated observational 
errors \citep{rajaguruetal07} and systematics in the analysis procedure
\citep{rajaguruetal06}. In this Letter, using waves with a first skip
travel distance sufficiently larger than the diameter of sunspot under study,
and employing deep-focus wave-path geometries that largely
avoid wavefield observed within the sunspot, we present new time-distance helioseismic
diagnostics that contrasts near-surface perturbations with deeper ones.
Signatures of a deep (up to about a depth of 18 Mm) increased wave speed
region, obtained using waves in deep-focus geometry are
presented in Section 3. Further support for this inference, from 
frequency dependences of surface- and deep-focus travel times, are presented in Section 4. 

\section{Data and Analysis Method}

A large sunspot (NOAA AR9057, diameter $\approx$ 32 Mm) that crossed the central meridian on
28 June, 2000 and observed by SOHO/MDI \citep{scherreretal95} in the full-disk resolution mode has
been chosen for this study. We extracted a data cube of size 373 Mm$^2$ by
1024 minute and have also obtained vector magnetograms
of the same region observed by IVM instrument at Hawaii (Hannah Schunker, private communication).
The data were pre-processed through the Stanford data pipeline system
to remap and track the region under study.

\begin{figure}
\epsscale{1.0}
\plotone{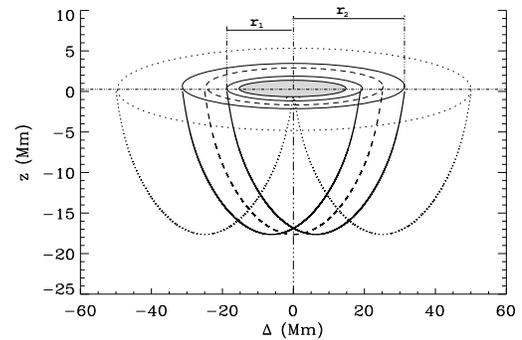}
\caption{Ray-path diagram depicting the surface- and deep-focus measurement geometries: the ray paths
correspond to model S of \citet{jcd-model}. Dotted lines correspond to the well-known 
center-annulus geometry, and solid and dashed lines correspond to annulus-annulus deep focus
geometry with q=0.6 and 1 (focus at the lower turning point), respectively}  
\label{fig:1}
\end{figure}

We choose waves having a first skip travel 
distance, $\Delta$, of 50 Mm that is larger than the diameter of the sunspot.
Apart from a phase speed filter that selects waves with horizontal phase speed around
54 km/s with a width (FWHM) of 21.25 km/s, Gaussian frequency filters
centered around 2.5, 3, 4 and 5 mHz with widths of 1 mHz are also used to study frequency dependences
of travel times \citep{braunandbirch06}.
We employ three distinct measurement schemes (Figure 1):
(i) the traditional center-annulus surface focus geometry (dotted lines),
which, for the travel distance chosen, makes use of wave-field within sunspot 
either as a 'source' (out-going waves) or as a 'receiver' (in-going waves) 
but not both simultaneously, (ii) double skip center-annulus surface focus geometry,
which completely avoids sunspot oscillation signals for measurements over a region of
radius 43 Mm from the spot center, and (iii) an annulus-annulus deep focus geometry
with focus depths, $z_{d}$, ranging from 10.24 - 17.65 Mm (solid and dashed lines correspond
to $z_{d}$ = 16.85 and 17.65 Mm, respectively; details in Section 2.1).

\begin{figure}
\epsscale{1.0}
\plotone{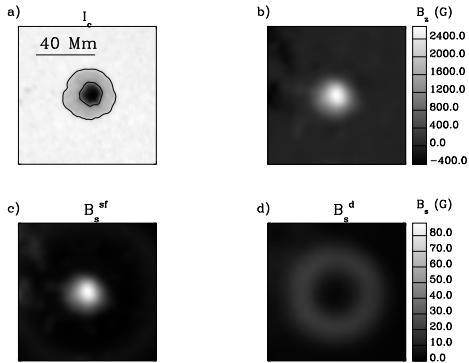}
\caption{a). MDI continuum intensity ($I_{c}$) image of the sunspot (contours 
mark the umbral and penumbral boundaries, b) smoothed, by a 3 pixel width Gaussian, 
IVM line of sight magnetogram rescaled to MDI resolution,
c) and d) the reduced magnetograms $B_{s}$ derived using Eqn.(1) for surface- and deep-focus
(average of q=0.75,0.8 and 1 cases) geometries; see text for more details}
\label{fig:2}
\end{figure}

Most studies have used the LOS magnetic field on the surface as a proxy for characterizing 
'surface signal' in the seismic measures \citep{lindseyandbraun05a,lindseyandbraun05b,
braunandbirch06}. We note that, in the presence of a deep perturbation
extending from the surface, surface magnetic field can be a proxy for neither the surface
effects nor deeper perturbations in seismic signals individually and exclusively.
Since a seismic quantity, either a wave travel time or phase, measured at any single location 
(pixel) uses wave field observed at distant points that are spread over an annulus or pupil,
a more appropriate proxy to characterize the surface effect would be similarly averaged 
surface magnetograms: weighted convolution of pupil functions, $a(x,y)$, (or masks for 'center'
and 'annulus' locations used to average the wave-field) with the magnetograms, $B(x,y)$, given by,
\begin{equation}
B_{s}(x,y)=\int a(x-x',y-y')B(x',y')w(r)dx'dy'
\end{equation}
that gives each pixel a weighted average of surface magnetic field over all the pixels from
where the wave-field used in a travel time measurement has come from (Figure 2).
Weights, $w(r)$, for magnetic field at a given location ($x',y'$) is chosen to be the inverse of
its horizontal distance, $r$, from the focus point (the measurement point,$x,y$), 
because the density of rays or wave-paths projected on to the surface go as inverse
distance from the focus. We use such reduced or averaged magnetograms, $B_{s}(x,y)$, 
to study any direct surface magnetic field induced variations in travel times.
The magnetic field $B(x,y)$ used here are LOS field strengths obtained from
IVM vector magnetograms rescaled to MDI full-disk resolution.

\subsection{Deep-focus measurement scheme}

Two concentric annuli, centered around the point of measurement, are used in our typical
deep-focus measurement scheme, similar to the original scheme proposed by \citep{duvalletal01}: 
a point from the inner annulus is cross-correlated with a diametrically opposite point in 
the outer annulus, and all such point-to-point correlations
spanning the whole annuli (360 degrees) are averaged. In this scheme, the travel distance 
$\Delta=r_{1}+r_{2}$, where $r_{1}$ and $r_{2}$ are the radii of inner and outer annuli,
respectively; and, different focus depths are achieved by varying the ratio $q=r_{1}/r_{2}$
from 0 to 1, while keeping $\Delta$ fixed; $q=0$ is the traditional surface focus, and
$q=1$ is deep focus at the lower turning point. This latter case measurements were first
performed by \citet{duvalletal01}. Here, we perform 8 different focus depth measurements
for the chosen $\Delta = 50$ Mm, including the surface focus; Table 1 provides focus
depths $z_{d}$, for the chosen $r_{1}$ and $r_{2}$, computed using ray theory for standard solar model
(model S of \citet{jcd-model}) and corresponding umbral averages $B_{su}=<B_{s}>_{umbra}$. 

\begin{deluxetable}{rrrr}
\tablecolumns{4}
\tablewidth{0pc}
\tablecaption{Annuli Radii, Focus Depths and $B_{su}$ ($\Delta$=50 Mm)}
\tablehead{
\colhead{$r_{1}$(Mm)} & \colhead{$r_{2}$(Mm)}   & \colhead{$z_{d}$(Mm)} & \colhead{$B_{su}$(G)}}
\startdata
0. & $\Delta$ & 0. & 60.0 \\
6.25 & 43.75 & 10.24 & 43.0 \\
10.0 & 40.0 & 13.12 & 29.0 \\
12.5 & 37.5 & 14.53 & 20.0 \\
18.75 & 31.25 & 16.85 & 6.5 \\
21.43 & 28.57 & 17.35 & 4.0 \\
22.22 & 27.78 & 17.45 & 3.7 \\
25.0 & 25.0 & 17.65 & 3.0 \\
\enddata
\end{deluxetable}

For deep focus measurements over the entire area of sunspot to be completley free of 
wave-field within the sunspot, both $r_{1}$ and $r_{2}$ should be greater than diameter
of the sunspot, i.e. $\Delta$ should be greater than 64 Mm; but changes in travel times
of waves with such large $\Delta$, as measured in local methods, are too small to analyse the
effects under study.
Here, with $\Delta = 50$ Mm, we aim at getting at least the measurements within umbral area
to be free of wave-field within the sunspot, i.e. $r_{1}$ and $r_{2}$ should at least be
the sum of umbral (5.8 Mm) and full sunspot radii (16 Mm). 
From Table 1, we see that the last 3 (deeper most) annuli combinations roughly satisfy
such a criterion (see also Figure 2).

\section{Separating surface effects: deep-focus measurements}

\begin{figure}
\epsscale{1.0}
\figurenum{3}
\plotone{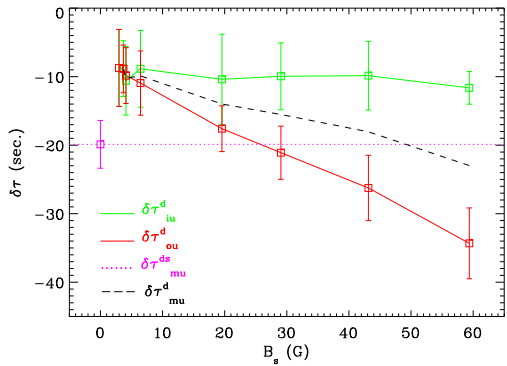}
\caption{Variation of umbral area averaged one way and mean travel times against $B_{su}$
for surface- and deep-focus geometries; see text for details.}
\label{fig:3}
\end{figure}

In deep focus geometry, contribution of oscillation signals from within the sunspot
to a given measurement, as characterized by $B_{s}$, decreases as the focus depth increases;
passage to $B_{s} \approx $0 G (Table 1 and Figure 2) occurs mainly in the
umbral region, and so we study the variation of umbral averaged travel 
times against, $B_{su}$, the umbral averaged $B_{s}$ (umbral area is marked 
by the inner contour in Figure 2a, and it is about 5.8 Mm in radius).
Figure 3 shows the variation of changes, with respect to a quiet Sun average, in umbral 
averages of one way (in- and out-going) and mean travel times, $\delta\tau_{iu}^{d}$, $\delta\tau_{ou}^{d}$ 
and $\delta\tau_{mu}^{d}$, against $B_{su}$.
The error bars correspond to standard deviations of travel times over the umbral pixels.
Surface-focus geometry yields a value of $B_{su}=60$ G, and the corresponding travel times 
$\delta\tau_{iu}^{s}$ and $\delta\tau_{ou}^{s}$ are the two right most data points in Figure 3.
Umbral average of half of double skip travel times, $\delta\tau_{mu}^{ds}$, 
which have only a mean signal and are completely free of wave field within the 
sunspot (cf. Section 2) corresponding to a zero $B_{s}$, is given by the dotted 
horizontal line. 

Several important pieces of information about the sub-surface structure
and interaction of p modes with the sunspot magnetic field are in order, in the variation of travel times
against $B_{su}$ shown in Figure 3, once we understand the variation of measurement geometry and
dependent sensitivities to travel times. Firstly, because of geometry and the
azimuthal averaging involved, the sensitivity to flows in deep focus measurements gradually 
decreases as $z_{d}$ increases and is zero when $z_{d}$ is the lower turning point for waves. This
happens independent of $B_{s}$. On the other hand, passage towards $B_{su}=$ 0 G involves wave-paths
changing from in and out of sunspot to that of crossing it at some depth.
Because of the above reasons, the asymmetric
variation of one way travel times to a non-zero mean value of $\delta\tau_{mu}^{d} \approx $ -10. sec
at $B_{su} \approx $0 G is accountable solely neither by material flow and its gradient in depth 
nor by effects localized near the surface due to predominant surface magnetic field interactions. 
An increased wave speed region extending down to, at least, a depth of about $z_{d}-\lambda_{h}/2$,
where $\lambda_{h}$ is the horizontal wavelength of waves with dominant power within the 
frequency band used, is essential here. 
For the model S of \citet{jcd-model} the asymptotic relation (ray theory)
$\lambda_{h}=R_{\odot}c(r_{t})/\nu r_{t}$ yields a value of about 15 Mm for frequency $\nu$=3.5 mHz and
lower turning point $r_{t}$=18 Mm, and so the increased wave
speed region should extend at least to within 7.5 Mm from the focus depth around 18 Mm. This
is broadly in agreement with early time-distance helioseismic inversions for sound speed
\citep{sashaduvall99}, as far as the depth extent of positive sound speed changes is concerned.
Secondly, attributing the largely $B_{su}$ independent (hence, focus depth independent) values
for $\delta\tau_{iu}^{d}$ ($\approx $ -10. sec.) solely to seismic signals from flow and sound speed
requires a following surprising coincidence: (down)flow and sound speed perturbations
both have negative depth gradients such that their signatures in $\delta\tau_{iu}^{d}$ (ingoing waves travelling
against the flow) exactly cancel out, while those in $\delta\tau_{ou}^{d}$ (outgoing waves travelling 
with the flow) add up leading to the changes seen in Figure 3. Frequency dependences of
$\delta\tau_{iu}^{s}$, $\delta\tau_{ou}^{s}$ and $\delta\tau_{mu}^{ds}$, shown and discussed in the next
Section, however, argue against such a scenario. Moreover, we find that half the 
double skip travel time $\delta\tau_{mu}^{ds}$ (-20. sec.) is closer to $\delta\tau_{iu}^{s}$ (-11. sec)
than to $\delta\tau_{ou}^{s}$ (-35. sec), confirming the early findings of \citet{braun97}. This
result, again, depends on wave frequency as shown in the next Section.
 
In summary, results in Figure 3 show that use of wave-field observed within magnetic regions 
with significant $B_{s}$ leads to travel time measurements that cannot be explained solely in 
terms of seismic signals due to flows and sound speed. However, our measurements in the
limit of $B_{su} \approx $0 G, where both surface magnetic and flow contributions go to zero,
show a significant mean travel time signal indicating a faster wave speed region extending
down to about 18 Mm with a resolution uncertainty of $\lambda_{h}/2 \approx $ 7.5 Mm. 

\section{Asymmetric frequency dependences of one way travel times}

Frequency dependence of mean travel times, over a sunspot region, was shown and interpreted by 
\citet{braunandbirch06} as an indication of perturbations largely confined to a region
not deeper than a few Mm. Here, we show in Figure 4 the umbral averages 
$\delta\tau_{iu}^{s}$, $\delta\tau_{ou}^{s}$, $\delta\tau_{mu}^{ds}$ and the deep focus times $\delta\tau_{mu}^{d} (B_{s}\approx 0)$ 
against frequency; $\delta\tau_{mu}^{d} (B_{s}\approx 0)$ shown here are the ones averaged over the three deep
most foci measurements (Table 1).

Two important results concerning the sunspot - p mode interactions and deep structure of sunspot
perturbations are contained in the frequency variations shown in Figure 4.
(1) There is a large asymmetry in the frequency dependences of $\delta\tau_{iu}^{s}$ and $\delta\tau_{ou}^{s}$:
dominant frequency dependence is seen only in $\delta\tau_{iu}^{s}$, while that in $\delta\tau_{ou}^{s}$ is very small,
for the chosen $\Delta$ of 50 Mm. A shallow sunspot perturbation cannot introduce such a direction
dependent frequency variation; moreover, $\delta\tau_{mu}^{d} (B_{s}\approx 0)$ are largely independent of frequency.
So, reasoning in the same lines as \citet{braunandbirch06} rather suggests a deeper extension for
wave-speed perturbation, provided we identify an independent physical cause for the frequency dependence
of $\delta\tau_{iu}^{s}$, which we do in case (2) below. We note that for $\Delta$ smaller than the 
diameter of sunspot, as is mostly the case in \citet{braunandbirch06}, and also from our 
results not discussed here, both $\delta\tau_{iu}^{s}$ and $\delta\tau_{ou}^{s}$ are equally frequency dependent.
The asymmetric frequency dependence at larger $\Delta$, then, is likely due to 'source' and 'receiver'
locations being outside or inside, but not both together within the spot: whenever waves from outside
the spot are 'received' inside, there is a large frequency dependence but no or little frequency 
dependence vice versa. 
\begin{figure}
\epsscale{1.0}
\figurenum{4}
\plotone{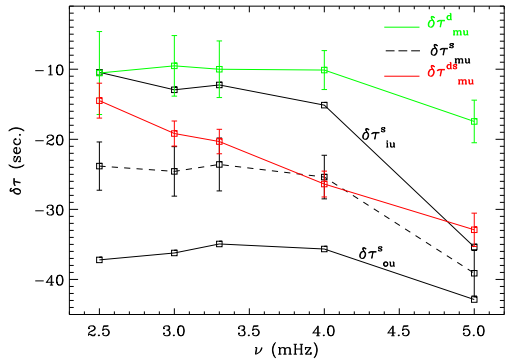}
\caption{Frequency dependences of surface- and deep-focus travel times (one way and mean).
Labels for different line types and colors identify the different measurements.
For clarity, only error bars for mean travel times in surface focus (dashed line) measurement
are drawn.}
\label{fig:4}
\end{figure}
(2) The mean times $\delta\tau_{mu}^{ds}$,
which correspond to waves with both 'sources' and 'recievers' being outside but focussing at a 
first skip surface location within the spot, are not only closer to $\delta\tau_{iu}^{s}$, but also have the same
frequency dependence. This points to an intimate connection, helioseismically, between ingoing waves
that survive the interaction with the sunspot to emerge on the other side and those 'observed' or 
'received' within the sunspot: we interpret this as a helioseismic signature of well observed
p mode absorption of sunspots. Further, as evident in Figure 4, the lower the frequency the smaller is
the difference between $\delta\tau_{mu}^{ds}$ and $\delta\tau_{iu}^{s}$. This might indicate that mode conversion
is more efficient at lower frequencies, for a given phase speed, than at higher frequencies or to a
wave-number dependent sunspot - p mode interaction.

\section{Discussions and summary}

Reliable three dimensional tomographic imaging of the whole depth range of sub-surface layers
of a sunspot is hard to achieve without having to use waves skipping at distances smaller than 
the size of a sunspot and hence without explicitly taking into account the magnetic field - p 
mode interactions. Since we lack a suitable model for such interactions and it has been
quite clear from various studies (\citet{lindseyetal07} and references therein) that significant seismic 
contributions arise due to surface magnetic field interactions, physical,
instrumental and systematic, most sunspot local seismic inversions have suffered from unreliable
depth descrimination of flow and sound speed structures. Here, we have attempted 
a completely heliosiesmic diagnostic approach to checking the extent of surface magnetic
effects by way of combining of surface- and deep-focus time-distance helioseismic
measurements that avoid oscillation signals observed within the sunspot.
Usian an appropriate surface magnetic field proxy, $B_{s}$, derived from 
magnetograms, we have been able to contrast the near-surface perturbations with deeper ones. 

Even though we have not proved that deep-focus measurements indeed have maximum
sensitivities localized enough at the foci, with wave propagation calculations
(which should form part of a future investigation), we believe that results presented
in Sections 3 and 4 together provide strong observational evidences for an extended
depth region of faster wave propagation beneath sunspots.
Large asymmetry of one way travel times of waves focussing near the surface within
a sunspot (i.e. when the surface magnetic field contribution is high) is also shown 
to be accompanied by asymmetric frequency dependence. We have identified such direction and 
surface magnetic field (or focus depth) dependent
changes in frequency dependence of travel times to be helioseismic signatures of 
p mode absorption and mode conversion. The surface magnetic proxy $B_{s}$ that we have
defined in Eqn.(1) depends only on how oscillation signals on the surface are used in a
given measurement and so should prove useful in characterizing 'surface magnetic' effects
in different methods of local helioseismology.


\acknowledgments
This work was supported by NASA grants NNG05GH14G to $SOHO$/MDI project and NNG05GM85G to Living With a Star 
(LWS) program at Stanford University.
The author thanks Dr. Sebastien Couvidat for a ray tracing computer code used in this work,
and Dr. Richard Wachter for discussions that helped understanding the results in this paper better.
Thanks are also due to Dr. Baba Verghese for help with Figure 1.


%
%
%
%
%
%

\end{document}